\documentclass[]{article}

\usepackage{arxiv}

\usepackage[utf8]{inputenc} 
\usepackage[T1]{fontenc}    
\usepackage{hyperref}       
\usepackage{url}            
\usepackage{booktabs}       
\usepackage{amsfonts}       
\usepackage{nicefrac}       
\usepackage{microtype}      
\usepackage{lipsum}		
\usepackage{graphicx}
\usepackage{natbib}

\usepackage{doi}

\begin{document}

\title{Beyond ``Fairness:'' Structural (In)justice Lenses on AI for Education}


\author{
    Michael Madaio\\
	Microsoft Research\\
	New York, NY \\
	\And
    Su Lin Blodgett \\
	Microsoft Research\\
	Montreal, Canada\\
	\AND
	Elijah Mayfield \\
	University of Pennsylvania \\
	Philadelphia, PA \\
	\And
	Ezekiel Dixon-Román \\
	University of Pennsylvania \\
	Philadelphia, PA \\
}

\date{}

\renewcommand{\shorttitle}{Beyond ``Fairness:'' Structural (In)justice Lenses on AI for Education}


\maketitle

\begin{abstract}
Educational technologies, and the systems of schooling in which they are deployed, enact particular ideologies about what is important to know and how learners should learn. As artificial intelligence technologies---in education and beyond---may contribute to inequitable outcomes for marginalized communities, various approaches have been developed to evaluate and mitigate the harmful impacts of AI. However, we argue in this paper that the dominant paradigm of evaluating fairness on the basis of performance disparities in AI models is inadequate for confronting the systemic inequities that educational AI systems (re)produce. We draw on a lens of structural injustice informed by critical theory and Black feminist scholarship to critically interrogate several widely-studied and widely-adopted categories of educational AI and explore how they are bound up in and reproduce historical legacies of structural injustice and inequity, regardless of the parity of their models' performance. We close with alternative visions for a more equitable future for educational AI.
\end{abstract}


\section{Introduction}
\label{section_1}
\setlength{\parindent}{2em}

During the COVID-19 pandemic, countries around the world closed their schools in successive waves, impacting nearly 1.25 billion students across all levels of education.\footnote{https://www.worldbank.org/en/data/interactive/2020/03/24/world-bank-education-and-covid-19}  
As institutions shifted to distance learning, or a hybrid of in-person and remote instruction \citep{reich2020remote}, educational leaders increasingly turned to technological solutions to provide continuity of learning for students \citep{teras2020post}. Administrators and teachers moved students into remote learning environments that may have never been chosen, if the world had stayed as it was in 2019. Students and the global education system will be impacted in enduring ways as the pandemic continues to exacerbate existing systemic inequities in education \citep{aguliera2020emergency,dorn2020covid}. 

As educational technologies are proposed as a solution for crisis-induced remote learning \citep{teras2020post,kizilcec2021mobile}, discussion of equity often centers around the assumption that increasing access to learning technologies will reduce inequity in educational opportunities \citep{hall2020pre,holmes2021ethics}. While some educational technologies may be beneficial for some learners \citep{kulik2016effectiveness}, the logic of distributive justice that equates access to technology as a form of equity \citep[e.g.,][]{holmes2021ethics,greene2021promise,pei2020attenuated} has motivated widespread adoption of learning technologies, as well as widespread data collection on computational traces of students' learning---data that is then used as part of educational algorithmic systems. Substantial research on data-driven algorithmic systems (often referred to as artificial intelligence, or AI\footnote{In this chapter, we use the term AI to refer to data-driven algorithmic assemblages broadly construed. Data-driven algorithms have a lengthy history in education, from computer-aided instructional systems in the 1970s \citep{carbonell1970ai}, to intelligent tutoring systems that provide feedback and instructions to learners on the basis of students' data \citep{pane2014effectiveness}, with more recent versions of such systems drawing on advances in machine learning \citep{piech2015deep} and natural language processing \citep{mayfield2019equity}. Following Richardson's writing on algorithmic decision systems \citep{richardson2021defining}, we use the term AI expansively, rather than narrowly scoped to refer only to systems that include technical components such as machine learning, given that many data-driven algorithmic systems may not strictly involve machine learning and are still often marketed under the label of---and conceptualized in the public consciousness as---AI, and moreover, data-driven assemblages are susceptible to---and further contribute to---the same biases pervasive in AI more generally.}) in other domains has demonstrated the potential for such systems to reproduce and amplify societal biases in a variety of ways (see \citet{barocas2016big} for an overview), or introduce novel sources of unfairness through disparities in their performance \citet[e.g.,][]{buolamwini18}, in domains as widespread as criminal justice \citep{richardson2019dirty}, healthcare \citep{obermeyer2019dissecting}, and hiring \citep{raghavan2020mitigating}. 
In spite of these known issues for AI systems, enthusiasm for AI in education has never been higher, with millions of dollars in funding for institutes of AI in education\footnote{https://www.colorado.edu/today/ai-education} and a recent National Science Foundation report on educational AI \citep{roschelle2020ai} that optimistically treats such systems as inherently beneficial, arguing that such systems may be solutions to educational inequity (cf. du Boulay, this volume).

In education research, quantitatively-oriented researchers may be motivated to use their training to evaluate biases in educational AI as a neatly quantifiable construct, with error bars and standard deviations, reporting differences in system performance or students' learning outcomes by demographic groups (e.g., race, gender, etc), as in Kizilcec and Lee (this volume). Indeed, this has been the dominant paradigm in the nascent field of algorithmic fairness (see \citep{hanna2020towards} for a critique). But social scientists and critical theorists of education have argued (e.g., \citet{dixon17,williamson2017decoding}) that such an approach to quantifying biases misses the bigger picture---that socio-historical processes both within and outside of the classroom shape the design, use, and evaluation of educational AI.

 In an educational crisis, it may be the case that concerns about equity are sublimated, in the interest of using any means necessary to continue business as usual within systems of schooling. Naomi Klein, writing about the response to Hurricane Katrina's impact on education in New Orleans, described how ``disaster capitalists'' such as Milton Friedman saw in that crisis an opportunity to ``radically reform the educational system'' of New Orleans through immense cuts to public education in favor of subsidies to for-profit private charter schools \citep{klein2007shock}. Friedman's response to the Katrina crisis was to redirect resources for public education towards for-profit schools. 
 
Yet, there are other responses to crisis. Social philosopher Pierre Bourdieu describes how moments of crisis present opportunities to challenge and shift dominant schemes of thought by ``rupturing'' existing social structures \citep{bourdieu1977outline}. The coronavirus pandemic is a global crisis in nearly every sector. Rather than further enabling technocapitalist interests in selling products that weave technology inextricably into the fabric of civic life\footnote{https://www.theguardian.com/news/2020/may/13/naomi-klein-how-big-tech-plans-to-profit-from-coronavirus-pandemic}, the pandemic could instead be used as inflection point for education decision-makers, researchers, and technologists to radically rethink the way that education is done and how our technologies are designed \cite{moore2021disaster}\footnote{https://www.de.ed.ac.uk/news/edtech-pandemic-shock-blog-ben-williamson-and-anna-hogan}. Given the known risks for (un)fairness and (in)equity in AI more generally, and recent enthusiasm about data-driven \textit{educational} algorithms specifically (or, educational AI), it is thus imperative that concerns about structural injustice are taken seriously, so that our educational AI systems do not simply reproduce legacies of inequity through their design. 

This chapter draws on the view from critical theory to interrogate and rethink what equity means for educational AI. We begin by summarizing the current state of research on algorithmic fairness, including approaches to measuring fairness of algorithms and their limitations at capturing more fundamental issues of structural injustice, informed by Black feminist theory. We then discuss the historical legacies of (un)fairness and (in)equity in systems of schooling more broadly\footnote{We focus in this chapter on the U.S. context, given the lived experiences of the authors. Further research should extend this lens to implications for educational AI in global contexts---as work such as \citep{sambasivan2021re,ismail2021ai,mohamed2020decolonial} have done for AI more generally.} and critically examine the operating logics of several widely-researched classes of educational AI systems. We uses these categories of educational AI, not to suggest that they are exhaustive, but as fertile sites to examine how educational AI may fundamentally perpetuate structural injustice, regardless of the quantitative ``fairness'' of the underlying algorithms' performance. Finally, we call for new methods and visions for justice-oriented approaches to research and design of learning science and educational technology, drawing on related calls for design justice, counterhegemony, and technology refusal in adjacent fields. We argue for the need for radical visions for the design (or dismantling) of educational AI in order to bring about a more equitable and justice-oriented educational future.

\section{Methods and Challenges to Measuring Fairness in AI} 
\label{section_2}

The rhetorical promise from designers of data-driven educational algorithms (i.e., educational AI) is that they are able to increase learning overall while reducing inequities, by helping all students learn in more personalized ways. One need only look at the theme of a recent AI in Education conference,\emph{ ``Education for All in the XXI Century,''} which solicited papers to explore how AI systems might contribute to reducing educational inequity. This line of thinking follows a broader tradition in technical fields that critics call \emph{techno-solutionism} \citep{morozov2013save}. Fundamentally, work in this vein assumes that technology can solve deep-seated, complex social issues or that technology is inherently less biased than humans. Critics have reason to be deeply suspicious of this theory of change. Before we explore those critiques further, though, we will first review prior work on measuring fairness in education, and then discuss the current state of research on defining and measuring (un)fairness in data-driven algorithmic systems, including machine learning and artificial intelligence (AI) and for such systems used in education. In other words, what \emph{does} it mean for a machine learning or AI system to be fair?

\subsection{Measuring fairness in educational assessments}
\label{education_fairness}
Statistical evaluations of equality for test questions across race and gender has a long history in the psychometrics of high-stakes exams like the SAT, going back at least six decades \citep{anastasi61,donlon81,sehmitt90}. In 1968, \citet{cleary68} studied the problem of fair test use across Black and white students for admissions to universities that had only just been racially integrated. Later, with ETS, \citet{linn73} undertook an extensive quantitative review of test subgroup fairness in the SAT, leaning on definitions from \citet{thorndike71}. From those early days to more modern research, the primary questions have long been the same: Are the psychometric properties of a test variant or invariant across groups? Is there differential performance for different groups, and, if so, does the test produce valid estimates for each group? Are the items on the test equally difficult for all students, holding their estimated abilities constant and, if so, are there systematic reasons for such differential item functioning? In more recent decades, Samuel Messick's and Lee Cronbach's concerns for the social consequences of test score use and interpretation have been given more attention. Older versions of these standardized tests famously included culturally discriminatory world knowledge like polo, regattas, ballet, and horseback riding \citep{weiss87}. The approach to solving this has been nearly uniform within the psychometric field: measure marginal distributions of outcomes by demographic group, and make use of items and tests that meet a set of defined criteria for group-level fairness \citep{hutchinson19}.

These research traditions have been focused on achieving reliable measurement and valid predictions of student ability across student populations. However, more fundamental questions have been sidelined from the mainstream discourse in the field---questions such as any interrogation of how ``ability'' was defined, or whether the basic prediction goal was desirable (or desired by the students), or who such an assessment might disproportionately harm, even if it was functioning ``correctly''  \citep{dixon17,Karabel06chosen,lemann00thebigtest}. These are thorny questions, and they are fundamental to interrogating the meaning of equity in educational technologies.

\subsection{Measuring fairness in AI and machine learning}
\label{ML_fairness}
Critiques of the social implications of technology are of course not limited to educational technologies. For decades, scholars such as Langdon Winner \citep{winner80do} and others \citep{winograd1986understanding} have argued that technology instantiates social and political relationships. Early work in this area examined several types of bias arising from computational systems \citep{friedman1996bias}, with other work critiquing the ``mythology'' of Big Data---``the widespread belief that large datasets offer a higher form of intelligence and knowledge that can generate insights that were previously impossible, with the aura of truth, objectivity, and accuracy'' \citep{boyd2012critical}---while other work questioned the ability of existing American anti-discrimination law to address algorithmic discrimination \citep{barocas2016big}. In spite of this rich variety of interdisciplinary critical scholarship, the field of AI and machine learning has recently converged on a narrower framing of how to understand the harms that AI systems might engender.

In recent years, academic approaches arising from the nascent field of fairness, accountability, transparency, and ethics in AI and machine learning (FAccT)\footnote{https://facctconference.org/} have primarily centered on evaluating the \emph{fairness} of algorithmic decision-making systems. Algorithmic decision systems are used in a variety of high-stakes domains to support, or in some cases replace, decisions about access to resources and opportunities, for applications as varied as healthcare services \citep{obermeyer2019dissecting}, parole sentencing \citep{angwin16,dixon2019algorithmic}, and student-school matching \citep{robertson2020if}. Current approaches to assessing biases in such algorithms have developed mathematical formulations of \emph{group fairness} \citep{kleinberg2016inherent,corbett18} and \emph{individual fairness} \citep{dwork2012fairness}, largely for algorithms that perform classification tasks (i.e., predicting one (or more) of a set of categorical labels, as in \citet{kotsiantis2007supervised}). 

Group fairness metrics for classifiers typically measure the equality of predictive performance across groups defined by protected attributes.\footnote{United States law specifically defines the following characteristics as protected: race, color, religion, sex and gender identity, pregnancy, national origin, age, disability, veteran status, and genetic information. Internationally, specific protections differ by jurisdiction.} By contrast, individual fairness approaches look at similar \emph{individuals}, defining similarity in a task-specific way that typically does not rely on demographic labels, and ask whether similar users of a given technological system can expect similarly allocated decisions, error rates, or system quality and effectiveness \citep{gillen18,kim18}.
Proposed solutions for mitigating unfairness have typically involved developing more ``diverse'' datasets (i.e., more data from ``under-represented'' social groups) and algorithmic mitigation approaches  \citep{agarwal2018reductions}, although there are reasons for wariness about both proposed solutions.

\subsection{Challenges to measuring algorithmic fairness}
Challenges to measuring algorithmic fairness abound. For instance, \citet{chouldechova17} showed that a number of common fairness metrics cannot be simultaneously satisfied (while population base rates are different), requiring an ultimately socio-political choice by practitioners of which fairness metric to use. These choices have serious consequences for any fairness evaluation, and belie the rhetoric that developing more fair systems is a straightforward task and that models can simply be ``de-biased'' \citep{bolukbasi16}. In addition, investigations of how multiply-marginalized individuals are affected by decision-making systems are critical, but scarce. \citet{buolamwini18}'s path-breaking examination of facial recognition systems’ performance at the intersections of race and gender---finding that such systems perform particularly poorly on images of Black women---represents a notable exception. 

Moreover, measuring (and attempting to mitigate) algorithmic (un)fairness also typically requires gathering demographic data in the first place, which may threaten the privacy of marginalized groups, who have historically been most vulnerable to data collection and surveillance \citep{eubanks2018automating}. Studying if a lending system is racially biased, for instance, often involves collecting customers' racial information, which can be a fraught process that requires categorization of identities into a small number of fixed labels and the collection of potentially sensitive demographic data. Data collection and privacy concerns are not new for education researcher. There is significant regulation in some countries designed to protect students' privacy, such as the Family and Educational Rights and Privacy Act (FERPA) in the US \citep{daggett2008ferpa}, although there is contention about its ability to account for modern data infrastructures \citep{zeide2015student}. There has also been substantive critical writing about student data privacy issues in learning analytics, including concerns about the storage, sharing, and afterlives of student data \citep{slade2013learning}. 

Finally, approaches to evaluating algorithmic fairness have mostly been developed for domains where decision-making systems allocate opportunities or resources---e.g., lending decisions, criminal sentencing, or university admissions decisions \citep{kleinberg2016inherent}; errors by such systems give rise to what has been called \emph{allocational} harms \citep{crawford2017trouble}. 
However, AI systems, including many systems used in education, often do more than simply allocate opportunities or resources. Recent scholarship is also identifying the ways in which group-based and individual fairness approaches may be inadequate for the social or dignitary harms that may also arise, which \citet{crawford2017trouble} has termed harms of \emph{representation}. For instance, the reproduction of racist characterizations arising from Google Photos' labeling Black faces as gorillas \citep{simonite18,noble18}, or the reproduction of gendered expectations by digital assistants who respond coyly to sexual harassment \citep{fessler2017alexa}. These represent fundamentally different kinds of harms than those described above, and may not be well-captured by existing group or individual fairness approaches.

\subsection{Measuring fairness in educational algorithms}

Researchers working on educational algorithms, recognizing that their systems are not exempt from the same threats to fairness as in AI systems more broadly, have begun to apply methods for evaluating the disparate impact of educational algorithms, predominantly applying group-based approaches to conceptualizing fairness. In their review of the state of algorithmic fairness in education, Kizilcec and Lee (2020) discuss existing approaches to assessing fairness in educational algorithms, including research evaluating how the accuracy of a system designed to detect students' affect (e.g., confusion, frustration, boredom) differed for groups of students in urban and rural schools \citep{ocumpaugh2014population}; research evaluating how the accuracy of a model designed to predict the likelihood of student dropout differed by the gender of the student \citep{gardner19}, and more \citep{doroudi2019fairer,lee2020evaluation,kizilcec2020algorithmic}. In spite of extant work that takes a more critical lens to understanding such phenomena -- such as work exploring the ethical implications of resource allocation based on automated prediction \citep{prinsloo14} or the implications of making decisions to de-prioritize individual support on the basis of group outcome measures \citep{scholes16} -- such critical work has been sidelined in favor of group-based fairness approaches, popular in the AI and machine learning community more broadly, as exemplified by Kizilcec and Lee (2020). The implicit assumption of much work on algorithmic fairness, made explicit in Kizilcec and Lee (2020) is that ``the innovation is beneficial in general'' and the work to be done is to ensure that the benefits are fairly distributed (a stance echoed in recent work by other educational AI researchers as well, as reported in \cite{holmes2021ethics}). Such an approach, however, fails to address the classes of representational harms previously described and fails to address the second-order consequences of data collection; and, more fundamentally, there is good reason to be skeptical that this optimistic assumption is in fact true (see Holstein and Doroudi, this volume, for more discussion of various frames \citep[cf.][]{selbst19} with which to understand ethics in educational AI).

\subsection{Structural critiques of group-based fairness approaches}
\label{fundamental}

As research on algorithmic fairness has developed, scholars (particularly from sociology and science and technology studies (STS)) have brought more fundamental challenges to the algorithmic fairness approaches described previously, critiquing the orientation of such approaches towards technical issues of models and datasets, rather than towards the larger social and historical context (e.g., institutions and social structures) of injustice that manifests in technical systems. More specifically, sociological (and Black feminist) scholarship has long recognized that social forces of racism, sexism, cis heteronormativity, ableism, settler colonialism, and so forth, have shaped the foundations of our institutions and social structures.\footnote{We write as researchers in the United States, but similar systemic forces shape social structures everywhere, in different ways.} Central to these analyses is the understanding that these oppressions are \emph{systems}, encompassing both unjust distributions of political power and economic resources as well as the ideologies and discourses that justify these distributions \citep{feagin2000racist,delgado2017critical}. These systems are also ordinary, reproduced by the ``routines, practices, and institutions that we rely on to do the world's work'' \citep{delgado2017critical}. Finally, they are interlocking; the systems of sexism, racism, and so forth do not operate independently, but are intersecting structuring forces. 

This last property is captured by two conceptual models developed by Black feminist scholars: \emph{intersectionality} and the \emph{matrix of domination}. Intersectionality is a term coined by Kimberl\'{e} Crenshaw as part of her groundbreaking analysis showing that anti-discrimination law---which addressed gender- and race-based discrimination separately---was inadequate to address the overlapping discrimination experienced by Black women \citep{crenshaw1989demarginalizing,crenshaw1990mapping}. In highlighting how Black women were marginalized both by the feminist movement and by the civil rights movement, Crenshaw's analyses illustrated the need to move beyond single-axis analyses of injustice. Patricia Hill Collins' ``matrix of domination,'' meanwhile, is a model that ``helps us think about how power, oppression, resistance, privilege, penalties, benefits, and harms are systematically distributed'' \citep{collins1990black} (see \citep{costanza2018design} for a discussion of the matrix of domination in technology). This scholarship suggests that technologies developed in a world whose institutions are thus structured will reflect the values and legacies of this social and historical context, as they are steeped in the ideologies and discourses that maintain these institutions and their interlocking systems of oppression. It is thus critical to analyze these technologies through the lens of these oppressive institutions, ideologies, and discourses. Indeed, emerging scholarship is beginning to identify the consequences of existing fairness approaches' failure to account for this broader social and historical context. We outline a few of them here.

First, in attempting to treat different demographic groups the same, \emph{group fairness}-based approaches ignore the reality that institutions and ideologies work to maintain profoundly unequal social orderings, with the result being that different groups are not treated the same in society. Hanna et al. describe this situation in their examination of race in algorithmic fairness \citep{hanna2020towards}:

\begin{quote}
   \emph{``By abstracting racial categories into a mathematically comparable form, group-based fairness criteria deny the hierarchical nature and the social, economic, and political complexity of the social groups under consideration... In short, group fairness approaches try to achieve sameness across groups without regard for the difference between the groups... This treats everyone the same from an algorithmic perspective without acknowledging that people are not treated the same."}
\end{quote}

Second, in ignoring the sociological complexity of race, gender, and other frequently used attributes, fairness measurements may be inappropriate or inconsistent; for example, Hanna et al. delineate the many ways in which race has been operationalized in measurements of algorithmic fairness, from interviewer-observed, appearance-based classifications to self-declared racial classifications \citep{hanna2020towards}. Fairness approaches also often measure with respect to a single attribute, ignoring the complexities of interlocking oppressions, thereby making it impossible to understand how people who are multiply marginalized experience technologies differently \citep{costanza2020design}.

To illustrate this disconnect, consider data-driven algorithms designed to detect toxic language in online social media. These systems are well-documented as treating African American Language (AAL) social media messages as more toxic than Mainstream U.S. English (MUSE) ones \citep{davidson2019racial}, and for this performance difference, the systems have been rightly critiqued from a group fairness perspective \citep{sap2019risk}. But, using the lens of structural injustice, researchers must also recognize the ideologies of language driving this outcome, which construct MUSE as acceptable and normal and AAL as ungrammatical or inappropriate, and acknowledge the impact of these systems, namely the reproduction of harmful patterns of linguistic stigmatization, discrimination, and disenfranchisement, which maintain hierarchies of language and race and naturalize unjust political and economic arrangements \citep{rosa2017unsettling}. These findings have clear implications for educational technologies (e.g., \citet{mayfield2019equity,loukina19}), which we describe in more depth in the following section.

More broadly, analyses that do not engage with technologies' social and historical context fail to recognize how that context drives decisions at many points in decision-making systems' development and deployment pipeline, from the choice of problem definition itself (e.g., ``what is the problem to be solved?'') to the choices of data (e.g., ``whose data is prioritized?'' or, more likely: ``whose data can be easily acquired?'') and choice of evaluation metrics \citep{passi2019problem}. Without examining this context, fairness approaches often treat biases as emerging incidentally, rather than as reflections of the institutions and ideologies structuring society \citep{benjamin2019race}, or, even when such structuring forces are acknowledged, as in Kizilcec and Lee (2020, p.4), they are sidelined in favor of group-based fairness evaluations. Such approaches address only individual undesirable features, models, and datasets rather than the larger ideologies, institutions, and practices that produce them \citep{hoffmann2019where}. These approaches also fail to recognize how technologies (re)produce social meanings (cf. \citet{winner80do}), thereby maintaining these institutions and ideologies. In this way, algorithmic biases are treated as disconnected from both their origins outside of technology and their impact on society.

Finally, in proposing solutions that are technical in nature---involving data collection, more sophisticated model development, etc.---algorithmic fairness approaches assume that continued technological development is the solution to unfairness in decision-making systems, rather than asking whether some systems should be fundamentally re-designed, or perhaps not built at all \citep{cifor2019feminist,green2018data,keyes2019bones,bennett2019what,hoffmann2019where,baumer2011implication,barocas2020not}. Thus, algorithmic fairness approaches maintain traditional power relations between technologists and the communities by privileging technologists in the decision-making processes about how technologies should be designed and deployed \citep{hoffmann2020terms,kalluri2020don}. In sum, the current state of the art in understanding differential harms of AI largely centers around evaluating group-based and individual fairness metrics. Given the historical legacy of group-based approaches to evaluating educational assessments, it is understandable why the educational AI field has similarly adopted such approaches. However, as we have outlined, conceptual models from critical theory and Black feminist scholarship suggest that a lens of structural injustice can help shed light on the fundamental equity issues posed by data-driven educational technologies (i.e.., educational AI).

\section{Injustice in Educational AI}
\label{section_equity}

 One approach to engaging with equity in AI systems in education---what appears to be the dominant view---might be to see them as inherently beneficial for education and to seek to make those systems accessible and accurate for as many learners as possible (e.g., \citep{holmes2021ethics,kizilcec2020algorithmic,hansen2015democratizing}). Indeed, if one adopted the group-based fairness approaches from AI research wholesale, one might simply seek to improve the relative performance of an educational AI system for different populations of learners \citep{madnani17,loukina19,lee2020evaluation,kizilcec2020algorithmic}. However, given the interlocking systems of oppression in a settler colonialist, patriarchal, cis heteronormative, ableist society such as the United States, we argue that educational AI technologies that simply reproduce the status quo under such a ``matrix of domination'' will be fundamentally unjust. 
 
 Thus, following other critical interventions in reviews of education research (e.g., \citep{kumashiro2000toward,north2006more}), in this section, we examine several widely-studied paradigmatic categories of educational technologies using lenses drawn from critical theory to identify how their operational logics may reproduce structural injustice in education, regardless of the parity of their algorithmic performance. We highlight these categories of educational AI systems, not as an exhaustive list, but as a demonstration of how lenses from critical theory may help shift the focus of the dominant paradigm of educational algorithmic fairness discourse towards a focus on structural injustice. 

\subsection{Legacies of educational systems of oppression}
\label{educational_equity}
 The policies, practices, and discourses that have shaped (and are shaped by) interlocking systems of oppression have influenced every part of the current educational landscape, from the nature of the curriculum, to the pedagogical methods used to teach it, to the systems of assessing what students know, to the technologies designed to teach and assess that curriculum in specific ways. 
 At the time of this writing, Black children in America are more than twice as likely as white children to attend high-poverty schools\footnote{More than sixty years after the U.S. Supreme Court declared school segregation unconstitutional in their 1954 \textit{Brown v. Board of Education} ruling, many American schools remain heavily segregated, and the legacy of historical racial oppression hovers over the entire American school system. In a recent Economic Policy Institute analysis of data from the National Assessment of Educational Progress, nearly seven in 10 (69\%) Black students attended a school where a majority of students were Black, Hispanic, Asian, or American Indian \citep{garcia2020schools}. In contrast, only one in eight (13\%) of white students attended a school where a majority of students were people of color \citep{garcia2020schools}. Black and Hispanic students are also substantially more likely than white students to not complete high school, with 79\% and 81\% graduation rates, respectively, compared to 89.1\% of white students \citep{EDFactsD17}, for 2017-18, the latest year on record. Students with disabilities reported the lowest rates of school completion, with only 67.1\% completing high school.
}, and such intertwined oppressions have led to worse educational outcomes for Black students in high-poverty schools \citep{garcia2020schools}. 

Educational policies and funding mechanisms, ostensibly motivated by wanting to support greater equity, have instead reproduced the status quo of historical legacies of structural injustice in education. For decades, government funding for education has been largely tied to schools' performance on standardized testing, tracing back to the 1983 publication of A Nation at Risk and through the measurement-focused era of No Child Left Behind.  The logic here is that schools must first assess students to understand gaps in performance between students from different backgrounds, in order to allocate resources to close those gaps --- a similar logic to the group-based approach to algorithmic fairness described earlier, with similar shortcomings. Indeed, performance on standardized tests has been shown to be highly correlated with parental wealth and families' ability to pay for expensive test-prep for their children---opportunities not often available to many marginalized learners \citep{dixon17}.


This assessment-focused paradigm has driven investment in and adoption of educational technologies to support these assessments, ostensibly to first measure and then mitigate gaps in students' performance. In the 2010s, American government funding for education was disbursed through the Race to the Top program, driving a surge in spending to move from summative to ongoing formative assessments, motivating increased adoption of educational technology in the classroom to deliver such ongoing assessments. But these investments in the early 2010s met numerous logistic hurdles, such as readily available broadband and wifi access, and the ongoing challenges of maintaining sufficiently updated computing devices \citep{boser12,fox12}. These gaps in access to educational technology infrastructures again fell disproportionately on already disadvantaged learners, further widening inequities in access to educational materials along socio-economic lines, which in the U.S. often means racialized communities \citep{warschauer10}. 

It is important to step back and critically question the assumption underpinning those investments --- that the issue at hand is simply providing access to the right technology that will reverse legacies of structural injustice impacting marginalized communities. This is a self-perpetuating logic, where group-level performance differences on assessments leads to investment and adoption of educational technologies in order to provide so-called personalized learning opportunities to reduce those performance differences for different groups of learners. In this ``distributive justice'' paradigm \citep{pei2020attenuated}, the goal is thus to ensure all students have equivalent access to those technologies (i.e., reducing group-level differences in access). 

With the transition to primarily online instruction during the COVID-19 pandemic, this paradigm became the dominant one. As schools around the world transitioned to remote learning during the crisis, school administrators, teachers, and parents look to technology to support ``continuity of learning'' \citep{reich2020remote,teras2020post,williamson20}. Much of the conversation has centered around how to provide equitable access to learning technologies, be that devices on which to learn, or access to digital learning platforms. Although providing continued access to educational opportunities in a crisis is certainly important, and ensuring equitable access to those learning opportunities is an important first step, it is critical to consider whether that enthusiastic adoption of technology has opened the door for data-driven educational AI algorithms that, once provided, will reproduce or amplify the structural injustices of the education system more broadly. 




To provide one brief example, consider classroom discussion practices. In her book \emph{Teaching to Transgress}, bell hooks argues that ``our ways of knowing are forged in history and relations of power'' \citep{hooks1994teaching}. Her classic example of this is the white male faculty member in a university English department who primarily teaches books by ``great white men'' and whose pedagogical practices perpetuate a white supremacist ideology. In many such classrooms, white male students may feel more comfortable volunteering in discussions more often, and may thus be called on more often by instructors \citep{hooks1994teaching}. When those discussions are mediated by technology---or moved entirely to a digital format, as in the now-ubiquitous remote learning exacerbated by the COVID-19 pandemic---those inequitable pedagogical practices are reproduced and amplified. One can imagine this playing out in a number of ways, from a technology that tracks students' participation in class or awards them points for such participation (perhaps inspired by work such as \citep{ahuja2019edusense}), to educational language technologies that summarize group meetings or class discussions (cf. \citep{li2019keep,zhao2019abstractive}). 

We thus extend hooks' argument to make the case that \textit{educational AI technologies} are forged in historical relations of power. 
Classroom pedagogical practices shape the design paradigms and goals of learning technologies, shape the data available for training educational AI systems, shape the ways in which those technologies are deployed in the classroom, and, more broadly, shape the values that are enacted by these systems (cf. \citep{winner80do}). Thus, the problem to solve is not simply providing \textit{more} students with access to such systems, or improving the accuracy of the systems' predictive models for different subgroups of learners (cf. \citep{kizilcec2020algorithmic}). Researchers and educational technologists must grapple with the legacies of historical disenfranchisement of marginalized learners and the impact those legacies have on the design paradigms of educational technology.

In the following sections, we will discuss a few such design paradigms of educational AI systems that reproduce historical legacies of educational inequity. We will discuss how a lens of structural injustice makes visible---and may allow us to challenge---the interlocking systems of oppression structuring the development and deployment of educational AI that a more narrowly scoped algorithmic fairness approach cannot. Through serious attention to these criticisms, we hope to generate provocative questions and directions toward alternative futures of educational AI technologies in the final section of the chapter.

\subsection{Biopolitical educational technologies}
\label{sel_biopolitics}
Here, we trace examples of categories of educational AI systems that may reproduce structural injustices --- regardless of the models' accuracy or fairness. We start first with biopolitical AI systems, or approaches that model students' socio-emotional states, in order to induce students to act in desired ways. Such socio-emotional systems are based on and perpetuate racialized ideologies about behaviors and emotions deemed appropriate or beneficial\footnote{Although prior work has identified affective states that are purportedly positively associated with learning \citep[e.g.,][]{craig2004affect,csikszentmihalyi1997flow}, recent work has critiqued the racialized construction of such affecive states in education \cite{zembylas2021sylvia}.} for learning \citep{williamson2017decoding}. In the Artificial Intelligence in Education community (AIED) and adjacent communities such as Educational Data Mining (EDM) and others, many papers have been published on the design and evaluation of systems that purport to model learners' ``affective state'' (e.g., \citep{henderson20194d,mandalapu2018towards,d2010multimodal}) and systems that use those estimates of learners' affective state to recommend specific types of content to learners (e.g., \citep{tsatsou2018adaptive,athanasiadis2017personalized}). These systems use audio, video, and other data sources to label students' ``affect,'' or emotional state as one of a set of pre-determined labels, as defined by the researchers, following from the broader field of affective computing \citep{picard2000affective}. Research in educational affect detection has largely converged around a set of (purportedly \citep[cf.][]{zembylas2021sylvia}) beneficial affective states, including boredom, confusion, and frustration \citep{d2010multimodal}, with some researchers also including ``engagement'' or ``flow'' \cite{csikszentmihalyi1997flow}, delight, and others \citep{okur2017behavioral,chang2018ensemble}.

The fundamental assumptions underlying these systems are that learners' affective states are legible to technology and that some affective states are more productive for learning than others \citep[e.g.,][]{craig2004affect,csikszentmihalyi1997flow}. One implication of this is that if AI models can identify such states, they can also `nudge' students into affective states that the systems' designers may believe to be more conducive for learning. Even if you believe that these affective states are associated with learning for all learners \citep[cf.][]{zembylas2021sylvia} (and setting aside unresolved questions of causality), the belief that one should thus `nudge' students \textit{into} these states represents an ideology akin to what Foucault and others have called \textit{biopolitics} or biopower. That is, a belief in the role of technology and policy to control people's biophysical states through detection, persuasion, coercion, or, when control is not possible, discipline \citep{williamson2017decoding,foucault1984right}. Biopolitical educational AI systems contribute to broader systems of algorithmic control through codifying the complex emotional states of learners in narrowly defined ways that are legible to the technologies, and which serve to reinforce particular, often racialized, ideologies about which emotions and behaviors are valued in the classroom, and which are punished \citep{gregory2017social}. 

Behavioral management and discipline in schools---particularly in America---is shaped by ideologies that reproduce structural injustices. School policies such as zero tolerance have led to racial disparities in student suspension and expulsion rates \citep{keleher2000racial,bottiani2017multilevel,hoffman2014zero}, with doubly marginalized students receiving greater disciplinary infractions \citep{garcia2020schools}. These racist disciplinary practices have contributed to the ``school-to-prison pipeline,'' where racialized students are disciplined at higher rates, suspended, expelled, and funneled into the prison system \citep{heitzeg2009education,kim2010school}. Structural injustice in schools' disciplinary practices is multifaceted, involving multiple, interlocking systems of oppression. In the GLSEN's 2017 National School Climate Survey, 
a majority of LGBTQ+ students (62\%) reported that their school had discriminatory policies or practices, such as disciplining public displays of affection disproportionately for LGBTQ+ students, with many reporting policies specifically targeting transgender students \citep{kosciw20182017}. 
Moreover, LGBTQ+ students with disabilities were more likely to be disciplined in school and to drop out of school than LGBTQ+ students without disabilities \citep{palmer2016educational}. Without considering structural injustice, the historical legacies of these policies and practices of schooling become baked into technological infrastructures underlying educational AI.

Let us take one educational technology company, ClassDojo, as an example. Despite plucky rhetoric around ``building classroom communities''\footnote{https://www.classdojo.com/}, ClassDojo was designed to allow teachers to track and reward students (via points) to incentivize what teachers feel is appropriate behavior in the classroom (and occasionally beyond) \citep{williamson2017decoding}. While ClassDojo is not currently a data-driven system in the same way as educational AI such as affect detection systems (which use data to train algorithms to identify affective states), the underlying logic of systems like ClassDojo arises from the same biopolitical ideology as affect detection AI and directly contributes to the creation of behavioral data that may be used to train biopolitical educational AI. Although a longer critique of biopolitical educational technologies such as ClassDojo is beyond the scope of this chapter (see \citep{williamson2017decoding}, or \citep{brynjarsdottir2012sustainably} for a critique of ``persuasive'' technologies more generally), ClassDojo is a clear example of how educational technologies may reinforce dominant ideologies of behavioral appropriateness---which construct racialized, LGBTQ+, disabled\footnote{We use the term ``disabled'' following scholars in the field of critical disability studies \citep{mankoff2010disability,spiel2020nothing,goodley2019provocations}.}, neurodivergent, and other marginalized students' behavior as inappropriate or criminal---instantiated into biopolitical technologies of behavioral discipline and control. Thus, the racialized disciplinary and behavioral practices and policies of schooling are reproduced and amplified by educational AI technologies that detect, track, and influence students' emotional states and behaviors in the classroom.


 During the ed tech boom driven by COVID-19 \citep{teras2020post}, educational institutions have increasingly\footnote{
While automated proctoring technologies (e.g., Proctorio, ProctorU, etc) had been in use prior to COVID-19, the pandemic accelerated adoption of these platforms, with Proctorio reportedly administering ``2.5 million exams --- a 900 percent increase from the same period last year''. https://www.vox.com/recode/2020/5/4/21241062/schools-cheating-proctorio-artificial-intelligence} turned to technology to extend their behavioral control into students' homes, with one egregious example being automated proctoring systems. Such systems encode particular ideologies of what is considered to be normal and abnormal (and thus suspicious) behavior during a remote test, ideologies that reflect and reproduce structural injustices in the education system more broadly. The algorithms that power such systems, including computer vision for facial detection and behavior detection algorithms, have been shown in other contexts to perform worse for women and people of color \citep{buolamwini18}. However, even if these models were to be tweaked in order to achieve parity of predictive performance across groups, the underlying assumptions undergirding these systems is that there is a set of normal---and thus desirable---behaviors, and behaviors other than these are suspected of being indicative of academic dishonesty. For neurodivergent learners, disabled learners, or learners with children or other caregiving responsibilities that might require them to behave in ways outside the narrow boundaries of what is considered to be normal behavior, the design of such systems reinforces larger systems of oppression that marginalize and punish any who fall outside of those encoded definitions of normality \citep{davis1995enforcing,swauger2020our,cahn2020snooping}. This ideology algorithmically encodes and perpetuates the carceral logics (i.e., ``cop shit''\footnote{https://jeffreymoro.com/blog/2020-02-13-against-cop-shit/}) of an education system that punishes marginalized learners in innumerable ways via the disciplinary practices of schooling.

\subsection{Educational surveillance technologies}
\label{surveillance}
Underlying and enabling these educational algorithms of biopolitical control---and for educational AI more generally---lies an ed tech surveillance infrastructure that collects data on students under the guise of care\footnote{https://www.insidehighered.com/blogs/college-ready-writing/surveillance-state}. Educational technologies---such as the aforementioned affect detection systems---often rely on or deploy a pervasive surveillance apparatus to collect multimodal (i.e., audio and video) data on learners' behavior, including in some cases from cameras deployed in classrooms to monitor students' posture, gestures, and movements \citep{ahuja2019edusense,martinez2020teacher,d2015multimodal}. Although not all educational surveillance approaches use AI, many draw on data-driven algorithmic methods to detecting the presence of people and their body movements (e.g., \citep{ahuja2019edusense}) or on language technologies for surveillance of students' communication (e.g., \citep{Underdig97:online}). In addition, educational surveillance assemblages provide the crucial infrastructure than ratifies and enables AI systems (as discussed in \citep{krafft2021action}). In a recent report from the University of Michigan's School of Public Policy, Galligan et al. argue that ``cameras in the classroom'' (or educational surveillance tools more broadly) reinforce systemic racism and discrimination, normalize surveillance for children, and further marginalize vulnerable groups.\footnote{\url{http://stpp.fordschool.umich.edu/sites/stpp.fordschool.umich.edu/files/file-assets/cameras_in_the_classroom_full_report.pdf}} 

Although some have argued that classroom surveillance systems can be used as a tool to help students or support teachers' professional development \citep{ogan2019reframing}, the reality of how these systems are marketed and used in schools around the world may belie this optimistic view, as classroom surveillance technologies are deployed often without students' awareness or consent and, once deployed, may be used for purposes other than they were intended for.\footnote{\url{https://www.sixthtone.com/news/1003759/camera-above-the-classroom}} In fact, although some tools for classroom monitoring appear to have been co-designed with input from teachers and students \citep{holstein2019designing}, the vast majority do not appear to incorporate student or teacher input. Even in such cases where students or teachers were involved in their design,
such approaches may merely be engaging in what has been referred to as ``participation-washing,'' where stakeholders' involvement in the design is used to ratify decisions already made by the system designers \citep{sloane2020participation,cooke2001participation}, a point to which we return in Section 4. Moreover, the normalization of cameras in the classroom, under the guise of care for students, has now, during COVID-19, led to the pervasive extension of surveillance into learners' homes via Zoom (or other video conference platforms used for learning) and the aforementioned automated proctoring systems.

In addition to cameras in classrooms and homes, as part of a pervasive state of surveillance, some school districts have deployed AI-driven language-based surveillance systems to monitor and extract text data from students' email and messaging platforms under the guise of ``school security'' (e.g., \citep{Underdig97:online}). These systems use natural language processing (NLP) models to flag messages or words that are hypothesized to be indicators that students will commit violent acts. The rhetoric of these platforms posits that schools that use them will be better able protect their students' safety, following a similar logic of paternalistic care as in other educational surveillance initiatives. However, educational technologies often act as ``racializing forces'' that encode and perpetuate racialized ideologies about language \citep{dixon2019racializing}, reinforcing the stigmatization and marginalization of minoritized language varieties and speakers \citep{rosa2017unsettling,blodgett2020language}. Consider the earlier discussion of how language technologies have been shown to identify speakers of African American Language on social media as more toxic than speakers of Mainstream US English \citep{davidson2019racial}.

Similarly, NLP systems have been shown to have higher false positive rates on text containing LGBTQ-related terms when attempting to detect hate speech \citep{dixon2018measuring}. On social media platforms, when these algorithms erroneously flag such language as hate speech, users may experience social or dignitary harms, where language about LGBTQ+ topics is stigmatized as inappropriate or offensive, and where (if such language is removed) their ability to participate in public discourse about these topics is diminished \citep{blodgett2020language}. This is bad enough. But in a school context, when a student has a message flagged as harmful or potentially violent, schools may dispatch social workers or disciplinary officers to intervene, thus (at a minimum) creating a disciplinary record for that student \citep{Underdig97:online}, or leading to serious harm for those students' lives and safety from the encounter with the carceral state. By creating disciplinary records and forcing interactions with institutions they might not otherwise have come in contact with, these surveillance systems directly construct ``misbehavior'' or ``criminality'' out of minoritized students' language use and may lead to their inclusion in databases of state apparatuses (cf. \citep{eubanks2018automating}).

This approach to widespread educational surveillance in the name of security---like the ubiquitous surveillance of the post-9/11 world---operates under what Brian Massumi has called the operative logic of pre-emption \citep{massumi2015ontopower}. These operative logics of mutual deterrence and pre-emptive strikes are not limited to war and statecraft, but infuse themselves into other aspects of our lives. Classroom surveillance technologies are one such platform, where a military logic of pre-emption informs the design and adoption of ubiquitous sensing infrastructure. 
Schools who purchase surveillance platforms such as the ``appearance recognition'' system Avigilon\footnote{\url{https://www.vox.com/recode/2020/1/25/21080749/surveillance-school-artificial-intelligence-facial-recognition}} or email and chat surveillance platforms Gaggle and Securly \citep{Underdig97:online} are able to argue that they have taken steps to make their schools safer from the supposedly ever-present threat of future school shootings, while having little evidence of their actual efficacy in pre-empting this future state. In the meantime, as surveillance infrastructure is adopted ever more widely, marginalized learners are the ones who are most harmed \citep{weinstein2020school}. Now, during the COVID-19 pandemic, similar logics of care and pre-emption have motivated school districts to contract with surveillance tech companies looking to capitalize on the crisis, leading to (among other surveillance efforts) proposals for students to wear badges to track their movements around schools\footnote{https://www.the74million.org/article/as-covid-creeps-into-schools-surveillance-tech-follows/}. The persistent legacy of racist school disciplinary practices suggests that these surveillance technologies will continue to reproduce structural injustice for marginalized learners.

\subsection{At-risk prediction technologies}
\label{risk_prediction}

The broad class of dropout prediction systems is another category of educational AI systems that reproduces structural injustice, regardless of the accuracy of such systems. Such systems focus on predicting learners' risk of failing to complete a class or a degree program, using behavioral data from interactions with a given course (e.g., \citep{tang2018time,nur2019student,lykourentzou2009dropout,liang2016machine,marquez2016early}). The underlying interventionist logic\footnote{Although there are, of course, other logics at work here, such as the business models that motivate much educational technology \citep[cf.][]{williamson20,teras2020post}.} of this line of research is that if researchers can identify which students are ``at-risk'' of dropping out of a course or a program, someone can intervene and prevent the student from dropping out. 
Decades of educational reformers have painted the ``dropout'' as a scary Other, designed to motivate learners through their fear of becoming part of this maligned class \citep{meyerhoff2019beyond}\footnote{These approaches follow what Eli Meyerhoff has described as a ``romantic narrative'' of educational progress---that learners advance through vertically arranged stages of educational progress, battling obstacles on the way to ascending to some imagined higher plane of existence after graduation \citep{meyerhoff2019beyond}.}. This Other has a racialized history. Educational researchers in the 1960s described minority learners as ``culturally deprived'' and argued that these students should be considered ``potential dropouts'' (with a too-brief caveat on the negative effect of such labels) \citep{hunt1966chapter}. Although the terms have changed, the underlying ideology of stigmatizing the ``dropout'' persists today, manifested in dropout risk prediction algorithms. 

Much like other research efforts for biopolitical educational technologies and surveillance systems that have found enthusiastic adoption in school systems, some universities are deploying what they describe as ``early warning systems'' that use predictive models to trigger an administrative action when a potential risk of dropout is detected, ostensibly in support of those students \citep{nam2019integrating}. This line of research on so-called ``Student Early Warning Systems''\footnote{\url{https://eliterate.us/instructure-dig-and-student-early-warning-systems/}}---evokes analogous risk modeling efforts in other domains, such as early warning systems for missile launches, disaster modeling, and financial risk models \citep{ahmad2013flood,lando2009credit,basch2000financial}. But there are consequences to viewing students as missiles or natural disasters\footnote{There is a longer tradition of pathologizing students as, for instance, patients in need of inoculation, as in \cite{ivan1973deschooling}}. Classifying or labeling learners as ``at-risk'' has historically had inequitable consequences for marginalized learners \citep{hunt1966chapter} in ways that become self-fulfilling prophecies, particularly when educators begin viewing students as at-risk. The language implicit in dropout ``risk'' prediction evokes the alarmism of the educational reform report, ``A Nation At Risk,'' which spurred panic about minoritized learners and led to decades of testing and policies ostensibly designed to help learners ``at risk.'' Further, the underlying data used to train these models is often a set of profoundly inequitable educational outcomes---the result of decades of racist educational practices---described at length in Section \ref{educational_equity} \citep{hooks1994teaching,giroux2001theory,keleher2000racial,kim2010school}. When such systems are used to predict college enrollment from students' behaviors in middle school \citep{pedro2013predicting}, these systems risk further entrenching legacies of historical injustice into future learning opportunities.

Beyond all of this, even if the models do successfully identify whether students may be likely to fail a course or dropout from a program, it is not clear how they may be used to \emph{prevent} students from leaving the course. One approach to this has been to use the output of the risk models to provide targeted advising to students, as in Purdue University's Course Signals program. Other universities have used such models to support what has been unsettlingly called ``intrusive advising'' or, providing proactive advising to students they think need it \citep{glennen1985reduction,varney2012proactive}. 
As in other AI systems in education, some researchers have begun to investigate the differential performance of these dropout prediction systems for different populations \citep{hutt2019evaluating}. Left unanswered (or largely unasked) by these algorithmic fairness approaches is the question of whether these models should exist at all, or the consequences for marginalized learners of being thus labeled.

During the COVID-19 pandemic, as schools moved online and many high-stakes assessments were canceled, several school systems decided to use algorithmic predictions of end-of-semester grades in lieu of assessments or teacher-assigned grades. Several high-profile examples of this\footnote{https://www.nytimes.com/2020/09/08/opinion/international-baccalaureate-algorithm-grades.html} highlight the risks of predictive modeling approaches to grades. In the UK, algorithmic predictions of students' A-level exams were found to be biased against students from schools in lower-income areas\footnote{https://www.theguardian.com/education/2020/aug/13/almost-40-of-english-students-have-a-level-results-downgraded}. Historical patterns of disinvestment in education and broader societal inequities were instantiated in the algorithm, reproducing these historical legacies of oppression. However, after many students saw their final predicted course grades deviated from what they expected, massive protests erupted, with rallying cries of ``F*ck the algorithm''\footnote{\url{https://unherd.com/2020/08/how-ofqual-failed-the-algorithm-test/}}. The UK Department of Education later canceled the algorithm, but it provides a cautionary tale for educational technologists who intend to develop predictive algorithms to model students' performance or risk (in part because the models may be re-purposed for other goals, such as grade assignment). Even if this algorithm had been assessed for group-based fairness \citep{kizilcec2020algorithmic}, the underlying premise of the algorithm was to assign grades on high-stakes assessments on the basis of historical performance on that assessment, reproducing systemic biases (cf. \citep{papageorge2020teacher}). 

Without a critical lens to question the historical legacy and sociopolitical context of scoring on assessments, these algorithms may perpetuate structural injustice. At-risk prediction algorithms thus continue to harm marginalized students (whether the racialized Other of students ``at risk'' or via socioeconomic inequities as in the UK A-level algorithm), in ways often left unquestioned in more optimistic views on educational technology. In this section, we have discussed three classes of educational AI systems --- biopolitical algorithms; the surveillance infrastructure that draws on and enables educational AI; and risk prediction. These categories of educational AI offer instructive sites with which to understand how educational AI may reproduce structural injustice in education, despite the best intentions of the researchers and designers, and in ways not accounted for by traditional group-based approaches to algorithmic fairness.

\section{Where Might We Go From Here?} 
\label{visions}

The question at this point is not \textit{whether} inequity exists in educational AI. The unequal treatment of students in today's educational settings is well-established, and we have argued that data-driven educational algorithms that fail to grapple with those socio-historical structures may reproduce and amplify existing inequities. Identifying group-based performance differences for various educational AI systems may be a good start, but it remains insufficient if those systems are built on---and further reproduce---existing systems of oppression, such as, for instance, racist language ideologies. However, individual researchers may feel unable to make significant changes to their research agenda (as suggested by related findings on industry AI research \citep{holstein19,madaio2020co,rakova2020responsible}), much of which has been built up over years and is hard to pivot. 
In the face of this, how might we as a research community re-imagine a better future, and confront the structural injustice of educational AI?

\subsection{The limits of representation and participation}

\label{limits}
One common response to these concerns is to acknowledge the risk of biases from technologies, and then shift to argue for their resolution through more diverse voices involved in the design and decision-making process. This push for representation as a solution is an important step towards addressing issues of access and opportunity to the knowledge, practices, and spaces that are shaping educational AI. As one example, there is a push to train more data scientists from underrepresented backgrounds; in addition to the inherent value of this, the argument is that if more diverse voices were part of the AI design process, their perspectives would be able to shape the design of algorithms\footnote{https://hbr.org/2020/10/to-build-less-biased-ai-hire-a-more-diverse-team}. 

While these efforts are laudable, as postmodern feminist and political philosopher Iris Marion Young argued \citep{young2011justice}, a focus on identity and inclusion is not enough to address the non-material processes of power and oppression. \citet{dixon2020data} argue that this approach is based on a politics of inclusion, which makes assumptions that diverse or representative participation will shift the epistemology and logic of socio-technical systems. However, \citet{dixon2020data} argue that this approach fails to address the normative disciplinary practices that produce knowledge and shape educational systems (cf. \citep{hoffmann2020terms}). Adding diversity to who is included at the table of decision making is an important initiative, but as Benjamin (2019) argues ``...\textit{so much of what is routine, reasonable, intuitive, and codified reproduces unjust social arrangements, without ever burning a cross to shine light on the problem}'' \citep{benjamin2019race}. \citet{dixon2020data} further argue that politics of inclusion fundamentally do not transform the norms and logic that make up the epistemology of the system, but may in fact maintain and reify the colonialist logic of hierarchies of difference that was formed in the post-Enlightenment and inherited by science and technology. In other words, the issue at hand is not just about representation in design, it is fundamentally about the epistemological \textit{hegemony} of educational AI --- that is, questions about what it means to produce knowledge, which kinds of knowledge are valued, and who determines what knowledge is valued in the design of technology. Thus, the idea of just adding (for instance) more diverse team members may not fundamentally shift or change the underlying ways of thinking and knowing and the structures of power that animate those ideologies. The recent example of Google firing the co-leads of their Ethical AI research group, Drs. Timnit Gebru and Margaret Mitchell \footnote{https://www.fastcompany.com/90608471/timnit-gebru-google-ai-ethics-equitable-tech-movement}, for raising concerns about the authenticity of Google's diversity and inclusion efforts, as well as their critical stance towards large-scale AI systems, throws into stark relief the limits of representation and inclusion of diverse voices in the research and design of AI systems, given existing power structures within technology companies.

An alternative theory of change looks to research programs that include more participatory design and community-based co-design research(ers)\footnote{https://participatoryml.github.io/}. This would ideally mean that research teams involve members of marginalized stakeholder groups as meaningful, co-equal members of the research team, or as advisors to the research projects, whose voices are given equal weight in framing the goals from the start, and who should be able to say that the research should not continue, if there is no way to make it equitable. This is in contrast to simply offering tokenized feedback to an already finished project, or a project that can only be altered at the margins (cf. \citep{arnstein1969ladder,sloane2020participation}). One can see such participatory approaches used in public policy \citep{corbett2018problem} and community-engaged health research \citep{balls2017use}, as well as in the tradition of participatory design and co-design in human-computer interaction \citep{harrington2019deconstructing,muller2007participatory}, which is emerging in the learning analytics field as well \citep{holstein2019designing}\footnote{http://pdlak.utscic.edu.au/call.html}.

While this is a positive step, much like calls for more diverse voices on AI development teams, there is reason to be skeptical of this type of ``reformist'' approach to change. Simply ``adding users and stirring'' \citep{muller2007participatory} will do little to change the underlying power structures shaping educational AI systems (see Holstein and Doroudi, this volume, for additional discussion of this point). Addressing structural injustice through participatory design will require reconsideration of what it means for stakeholders to participate in AI research. Researchers must ensure that their work does not further reproduce existing power dynamics and hierarchies in the broader society, both within the research teams and with their relationships to stakeholders \citep{sloane2020participation,cooke2001participation}. Such work will also need to avoid overburdening marginalized communities, by compensating them adequately for the full scope of their newly expanded participation and avoiding what some have referred to as the ``epistemic burden'' of extractive forms of participatory design \citep{pierre2021getting}. Truly empowered participation by marginalized groups in these research agendas requires the ability to change ways of knowing and what evidence looks like. It will require a willingness from the researchers to change the definition of efficacy and success in an intervention, based on what communities want and need. It will require looking at different disciplines and paradigms, beyond techno-solutionist paradigms and beyond well-tread fields like psychometrics and developmental psychology (and yes, algorithmic ``fairness''), and instead towards the voices of community organizers, social justice scholars, critical theorists and others. This will require a humility on the part of researchers and funders, a willingness to be part of a change that may leave them with less control at the end of the funding and research process than they had at the start.

The field of AI ethics more broadly is grappling with this issue as well. Despite dozens of values statements for ethical AI produced by large technology companies and government agencies \citep{jobin2019global}, 
AI systems continue to be developed that discriminate and perpetuate injustice. While there may be legitimate organizational reasons why AI practitioners have been unable to put these principles into practice (e.g., \citep{holstein19,Stark:2019gq,madaio2020co,rakova2020responsible}), others have been skeptical of these companies' intentions. Some critics describe these AI ethical statements as ``fairwashing'' or ``ethics-washing''--- that is, developing principles for ethical AI, while doing little to change the nature of such technology, or the organizational processes that led to their design \citep{bietti2020ethics}, acting as a ``smokescreen'' for business as usual \citep{sloane2019inequality}. Thus, while it may be worthwhile to begin with efforts to support more diverse perspectives on educational AI teams and more participation from marginalized communities in those systems' design, this will not be enough to transform fundamentally unjust systems.

\subsection{Equitable futures for AI in education}
As we close, we ask what an alternative vision for a more equitable future for educational AI might entail. First, and most critically, we argue that educational AI researchers and designers should adopt a \textit{design justice} approach to educational AI research. As proposed by Sasha Costanza-Chock, design justice is a ``framework for analysis of how design distributes benefits and burdens between various groups of people'' \citep{costanza2020design}. This involves interrogating the values encoded into designed systems, meaningfully involving members of marginalized and impacted communities in the design of technology, and, more generally, critically interrogating the narratives, sites, and pedagogies around design---here, the design of educational AI systems. We have begun such efforts in this chapter by tracing socio-historical legacies of educational inequity through several categories of educational AI systems that may perpetuate unjust ideologies and structures of oppression, drawing on lenses from critical theory and structural justice to inform this work.
To design more equitable educational AI, though, will require the learning science and technology community to learn, adopt, and promote theories and methods fields outside of what is considered to be a core research area for AI; approaches that may draw on design justice, critical theory, participatory design, and other areas. 

But changing the direction and goals of a research community means not only that individuals should change their research and design practices, but that the field as a whole should interrogate and confront the ways in which systemic educational inequities may be reproduced in the design, research, and policy of educational AI. What does it mean to be a justice-oriented learning science researcher? How might the existing incentives of the field be changed to encourage researchers to interrogate ethical issues and center justice-oriented decisions in their research practice? These incentives may include the grants and funding that learning scientists pursue, the collaborators they choose, the role of stakeholders in their research (including grappling with power hierarchies between and within communities of stakeholders), and researchers' ongoing engagement with the impact and potential inequities of their research after dissemination and deployment. Ideally, such an approach would entail a longer-term view of how research and development of educational AI (and educational technology more broadly) impacts learners' educational opportunities over the current, more incremental approaches. To what extent might values of justice and equity be adopted in conference and journal calls, review processes, and paper awards? In tenure and promotion reviews at universities? How might we avoid the co-optation of calls for equity that become neutered and narrowly scoped into neat technical fixes and ethics-washing? 

We, along with others in AI more broadly (e.g., \cite{benjamin2019captivating}), argue that the field of learning science and educational AI needs new, more radical visions for equitable learning futures. Visions for the kinds of technologies that can be built are shaped by what has been called ``\textit{socio-technical imaginaries}'', or: 
\begin{quote}
    \textit{The collectively held, institutionally stabilized, and publicly performed visions of desirable futures, animated by shared understandings of forms of social life and social order attainable through, and supportive of, advances in science and technology} \citep{jasanoff2015dreamscapes}.
\end{quote} The socio-technical imaginaries of our current field of educational AI are shaped and circumscribed by the historical legacy of the matrix of domination in education and technology. But they can be remade. To do this, we are inspired by Keyes et al.’s call for \textit{counterpower}, or what they call emancipatory autonomy in human-computer interaction, or, more simply, anarchist HCI \citep{keyes2019human}. For them, counterpower involves fostering community-appropriate and community-determined research and design, both between researchers and stakeholders as well as within the research community itself (here, learning science and educational AI), akin to what Asad et al. have referred to as ``academic accomplices'', or scholars whose research is designed to support (rather than displace) the already ongoing justice work on communities \citep{asad2019academic}. This move towards counterpower would involve everyone, not simply a privileged few, in taking control over the means to shape the forms of socio-technical educational systems\footnote{One can see echoes of calls for liberatory approaches to technology more generally \citep[e.g.,][]{benjamin2019captivating,winchester2018afrofuturism,tierney2019dismantlings}, and tensions in political traditions between participation and paternalism \citep[e.g.,][]{cooke2001participation,moir2013design}.}. What might such counterpower look like for learning science research, and how might it lead to new socio-technical imaginaries for educational AI?

For some research directions, it may be the case that the research should simply cease --- where ``the implication is \textit{not} to design'' \citep{baumer2011implication,barocas2020not}. Os Keyes poses a similar challenge by arguing that data science's approach to quantification and classification is intrinsically at odds with --- and a danger to --- queer lives \citep{keyes2019counting}. Predictive policing presents another such case, where generations of racist policing practices means that algorithms trained on these data will inevitably reproduce and exacerbate these practices \citep{dixon2019algorithmic,richardson2019dirty}. In such cases, injustices are inextricably embedded in the proposed technical solutions. For these types of tasks or methods, no amount of group fairness measurement, or UX optimization of a workflow will produce anything other than a reproduction and entrenchment of existing structures of oppression. Thus, the only recourse may be for people to organize to shut it down. In AI more broadly, public acts of resistance or refusal are becoming more common. For instance, communities have organized against computer vision used in public housing projects \citep{Voicesof50:online}, and the Stop LAPD Spying Coalition has successfully organized to ban predictive policing in Los Angeles, among other cities\footnote{\url{https://capp-pgh.com/}}. In education, one might look to the recent protests in the UK to ``F*ck the algorithm'' that predicted their A-level scores\footnote{https://unherd.com/2020/08/how-ofqual-failed-the-algorithm-test/} or organized resistance against automated proctoring systems \citep{cahn2020snooping}\footnote{https://www.eff.org/deeplinks/2020/09/students-are-pushing-back-against-proctoring-surveillance-apps}, among others\footnote{https://www.donnalanclos.com/listening-to-refusal-opening-keynote-for-aptconf-2019/}. 
However, these examples are acts of resistance and refusal to systems that are already designed, deployed, and causing harm in the world. 

We thus call for a wider range of design methods and theories to envision more radical, equitable futures for the learning sciences and AI in education. The design of our modes of learning is too important to be left in the hands of a privileged few. Current rhetoric from AI corporations and researchers \citep{moreau2019paradigm} is full of calls to  ``democratize''  AI, which mostly seems to refer to producing open-source tools for developing machine learning models (e.g., TensorFlow). In practice however, while platforms like TensorFlow may be freely available, simply giving more (often already privileged \citep{west2019discriminating}) people access to them will do little to change the fundamentally inequitable systems and power structures underlying the social contexts in which they are developed and used. Indeed, many current calls for ethical approaches to educational AI lean heavily on modeling after existing pedagogical paradigms (cf. du Bouley, this volume), which, as we discussed in section 3, may reproduce or exacerbate existing structures of oppression in schooling.

Instead, a more radical vision for democratizing data-driven learning technologies might reflect Paulo Freire's vision of a \emph{liberatory} pedagogy that allows learners to propose and engage with the problems that are essential for their lives \citep{freire70}. There is a rich lineage of resistance to centralized hierarchies in education, including the critical theorist Ivan Illich, who famously called for ``de-schooling society'' \citep{ivan1973deschooling} among others \citep{giroux2001theory,freire70,hooks1994teaching,meyerhoff2019beyond}. A full discussion of how the vision of Freire and Illich might be enacted in data-driven educational technology is beyond our scope here, but we are encouraged by recent work on the implications of a Freireian pedagogy for learning analytics \citep{broughan2020re}. Such traditions suggest additional opportunities for learners to take control and ownership over educational AI systems.

The design approaches and theories that fit neatly into the existing body of learning science and technology research have resulted in reproductions of existing, inequitable brick-and-mortar education systems. We have built algorithms that reflect back the inequities that are already in place. At best, educational AI researchers have conducted group-fairness comparisons for learners from different populations (e.g., \citep{kizilcec2020algorithmic,gardner19}). While such efforts are promising steps forward to mitigating fairness harms of existing systems, this retrenchment of existing algorithmic fairness paradigms will not lead to the creation of new socio-technical imaginaries, or to the design of more liberatory forms of education. As a field, we might instead look to methods from feminist speculative design \citep{martins2014privilege} and critical design \citep{bardzell2012critical,bardzell2013critical}, which endeavor to provoke and problematize \citep{forlano2014design}, and to have bold visions for designing more liberatory futures. 

One source for these new socio-technical imaginaries may come from queer theory. Scholars in queer theory have used the term ``queer'' as a verb, to describe a resistance to and deconstruction of binary and hierarchical categories and structures of oppression \citep{butler2011gender,ahmed2006queer,munoz2019cruising}. Although it originated in resisting rigid categories of gender and sexuality, it has been used to problematize hierarchical relations of power more broadly, including in the field of HCI and technology design \citep{light2011hci} and data science \citep{keyes2019counting}. Inspired by this, we ask what it would take to \textit{queer} learning technologies---to offer resistance to the dominant modes of thinking that reinforce systems of oppression in the learning sciences and educational AI. This might involve using critical design methods to provoke and problematize foundational assumptions in learning technologies, including what should be learned and what it means to have learned it -- to challenge dominant ideologies of what a ``good'' learner is and resist the technologies that treat deviation from the norm as suspicious.

Additionally, Dixon-Rom\'{a}n and others have argued that we might look to the field of Afrofuturism as another source for methods and theories to envision new socio-technical imaginaries and develop new visions for more liberatory futures \citep{de2017computational,benjamin2019captivating}. That is to say, if particular sociopolitical processes of colonial and post-colonial violence never occurred, how might socio-technical systems be designed differently, have different purposes, and enable alternative forms of reasoning other-wise? Afrofuturism entails a critical examination of the historical conditions that led to current societal inequities, while opening up a space for re-imagining alternative futures by asking the 'what if' questions that center an affirmative perspective of people of color \citep{dixon2019racializing, gaskins2019techno}. \citet{winchester2018afrofuturism} has described how Afrofuturism might ``place the often disenfranchised black voice central in the design narrative'' and ``plug the imagination gap'' in technology design. For a vision of how Afrofuturism might be used to counteract the ``racializing forces'' of educational AI, see \citet{dixon2019racializing}.

All of this may require rethinking the standard AI research and design lifecycle to prioritize equity and justice and to involve (and empower) stakeholders from marginalized communities throughout this lifecycle, in ways that avoid reproducing existing power dynamics in their participatory design approaches. This may entail a radical transformation of the ``institutionally stabilized'' structures and incentives of the current funding and research landscape to create welcoming spaces for diverse voices to shape the socio-technical imaginaries of educational AI and ed tech more broadly \citep{tuck2018toward,jasanoff2015dreamscapes}. It may require teaching learning science and educational AI students about the history and legacies of educational injustice, critical theory, and broader societal systems of oppression.

\subsection{Conclusion}
Educational policies and practices present visions of a particular type of learner, and attempt to prepare that learner for life in society---educational AI systems then instantiate these visions and values into algorithmic systems that may further entrench them. Whose visions are these? What type of learner is imagined, and what type of society are they being educated for? These are questions with critical consequences in a settler colonialist, white supremacist, patriarchal, cis heteronormative, ableist society. Paulo Freire and bell hooks, among other theorists and critics, have called for a more radical, liberatory form of education, where, rather than simply reproducing inequitable societal structures, the role of education is instead a provocation to rethink what is possible --- to remake the world in a more radical, more just, and more equitable way. 

This chapter interrogates issues of structural injustice in educational AI technologies through applying critical lenses to examining several forms of educational AI systems. The problems are complex and interwoven and resist straightforward analyses and answers---including quick technical solutions and neat group-level evaluations of ``AI fairness''. What is needed are approaches to transforming the epistemology of educational AI, working towards justice and equity---both in educational AI technologies and in the socio-technical and socio-political milieu in which they are designed. We do not offer solutions to the core challenges we have presented. To offer neat solutions would be to undermine the argument we have been making. 

What this historical moment, the crisis of the COVID-19 pandemic, has revealed is the potential for doing work outside the norms. We are in the midst of rapid and fundamental change to our educational system. We can use this as an opportunity to radically reimagine what we want for our education systems---including educational algorithmic systems and sociotechnical systems more broadly. Funding organizations and technology corporations are often the ones with the power and privilege to drive large-scale change. And yet as a community of researchers and practitioners, there is much we can collectively do to transform the trajectory of the field. There is an enormous rethinking about the role of technology in education that is beginning to occur worldwide. We must take this as an opportunity to reshape the priorities of the educational research we do and the research we support, towards addressing structural injustice in educational AI.

\bibliographystyle{plainnat} 

\bibliography{bibliography}

\end{document}